\documentclass{llncs}

\usepackage[utf8]{inputenc}
\usepackage[T1]{fontenc}

\usepackage[english]{babel}
\usepackage{csquotes}

\usepackage{amsmath}
\usepackage{amssymb}
\usepackage{enumitem}
\usepackage{orcidlink}

\usepackage{graphicx}

\usepackage{algorithm}
\usepackage{algpseudocode}

\usepackage{booktabs}
\usepackage{array}
\usepackage{multirow}

\usepackage{hyperref}
\hypersetup{
  colorlinks=true,
  allcolors=blue,
}

\usepackage[nameinlink,capitalize,noabbrev]{cleveref}
\makeatletter
\AddToHook{cmd/appendix/before}{\crefalias{section}{appendix}}
\AddToHook{env/algorithmic/before}{\def\@currentcounter{ALG@line}}
\makeatother

\crefalias{ALG@line}{line}

\usepackage[backend=biber]{biblatex}
\DeclareSourcemap{
  \maps[datatype=bibtex]{
    \map{
      \step[fieldsource=doi,final]
      \step[fieldset=url,null]
      \step[fieldset=isbn,null]
      \step[fieldset=issn,null]
    }
    \map{
      \step[fieldsource=eprint,final]
      \step[fieldset=url,null]
    }
  }
}

\usepackage{stmaryrd}
\newcommand{\simpreorder}{\stackrel{\subset}{\scriptscriptstyle\shortrightarrow}}
\usepackage[misc]{ifsym} 

\usepackage[acronym,shortcuts]{glossaries-extra}
\usepackage{glossaries}
\glssetcategoryattribute{acronym}{nohyperfirst}{true}

\setabbreviationstyle[acronym]{long-em-short-em}

\newacronym[longplural={deterministic finite automata}]{dfa}{DFA}{deterministic finite automaton}
\newacronym[longplural={finite automata}]{fa}{FA}{finite automaton}
\newacronym[longplural={nondeterministic finite automata}]{nfa}{NFA}{nondeterministic finite automaton}

\begin{document}

\title{Deconstructing Subset Construction
\thanks{This work will appear in TACAS 2026,  Lecture Notes in Computer Science (Springer).}
}
\subtitle{Reducing While Determinizing}

\author{John Nicol\inst{1}\textsuperscript{(\Letter)}\orcidlink{0009-0000-4060-7307} \and
Markus Frohme\inst{2}\orcidlink{0000-0001-6520-2410}}

\institute{
Unaffiliated, United States\\
\email{jnicol1994@alumni.cmu.edu}
  \and
TU Dortmund University, Germany\\
\email{markus.frohme@tu-dortmund.de}}

\maketitle

\begin{abstract}
  We present a novel perspective on the NFA canonization problem, which introduces intermediate minimization steps to reduce the exploration space on-the-fly.
  Central to our approach are \emph{equivalence registries} which track and unify language-equivalent states, and allow for additional optimizations such as convexity closures and simulation.
  Due to the generality of our approach, these concepts can be embedded in classic subset construction or \citeauthor{brzozowski1962canonical}'s approach.
  We evaluate our approach on a set of synthetic and real-world examples from automatic sequences and observe that we are able to improve especially worst-case scenarios.
  We provide an open-source library implementing our approach.
\end{abstract}

\section{Introduction}
\label{sec:introduction}

Finite-state machines are foundational models of computation in computer science, where their simplicity and expressive power make them of both theoretical and practical interest.
Among the most prominent types are \acp{dfa} and \acp{nfa}, which are extensively studied for their applications in various domains.
\acp{dfa} are often preferred over \acp{nfa} due to their efficiency in operations such as testing language equivalence and inclusion, universality, complementation, and other closure properties~\cite{HolzerK11,hopcroft2001introduction}, trace inclusion (noted in~\cite{van2008five}), and string matching~\cite{aho1975efficient}.
Many of these operations play a critical role in practical applications such as model-based testing~\cite{LeeY96,BrJKLP2005,UttingPL12} and theoretical fields such as automatic sequences~\cite{bell2020additive,shallit2022logical}.

\Citeauthor{RabinS59}~\cite{RabinS59} have shown that \acp{dfa} and \acp{nfa} are equally expressive by transforming \acp{nfa} into language-equivalent \acp{dfa} via \emph{subset} (or \emph{powerset}) \emph{construction}.
The resulting \acp{dfa} may be of exponential size compared to the original \acp{nfa} and may not be minimal.
The latter can be easily addressed via an additional minimization step~\cite{hopcroft1971n} which is often required if the resulting \acp{dfa} need to be processed further.
A minimal \ac{dfa} is called \emph{canonical} since a minimal \ac{dfa} is unique up to isomorphism~\cite{hopcroft2001introduction}.
Constructing the canonical \ac{dfa} for a given \ac{nfa} is called the \emph{canonization problem} \Cite{van2008five,tabakov2005experimental}.

The exponential blowup of canonization can be a challenge in practice.
However, many \acp{nfa} exhibit a notable \emph{compression ratio}, i.e., the ratio between the size of the determinized (via subset construction) \ac{dfa} and its minimized version.
Real-world cases report compression ratios ranging from $30$ (\cite{fici2022properties}), to $2500$ (\cite{Slavicki2021}), to $5$ million (\cite{Shallit2025b, Shallit2021}, and addenda for~\cite{mousavi2015mechanical}).
In such situations, materializing the fully determinized automaton is often the bottleneck of the canonization process.

For certain \ac{nfa}-related problems, there exists extensive research on limiting the impact of state blowup, e.g., by using \emph{antichains} for testing universality and language inclusion~\cite{doyen2010antichain,de2006antichains} or \emph{bisimulation up to congruence} for testing language equivalence~\cite{bonchi2013checking}.
See~\cite{chocholaty2024mata, garavel2022equivalence} for surveys on the current state of this work.
For the canonization problem, however, the construction of the full \ac{dfa} is necessary.

In this paper, we present OTF -- an \emph{on-the-fly} approach for canonizing \acp{nfa}.
Instead of treating determinization and minimization as distinct phases of the canonization process, OTF interleaves these concepts by repeatedly minimizing the (partially) constructed \ac{dfa} of subset construction.
Insights about equivalent states of the \ac{dfa} under construction can be used to detect equivalences even for yet unseen subsets of \ac{nfa} states and are fed back to subset construction in order to reduce the state space requiring exploration.
This form of communication is formalized via \emph{equivalence registries} with which the process can be parameterized, giving OTF a framework character.
We present two exemplary implementations of equivalence registries.
In our evaluation, we show that OTF is able to improve the performance of canonization for complex real-world \acp{nfa}, while remaining competitive for simple(r) systems.
We implement our approach in an open-source library~\cite{otf} that is based on AutomataLib~\cite{automatalib} and included in the theorem prover Walnut~\cite{walnut} since version 7.

A noteworthy alternative to classic \ac{nfa} canonization is \citeauthor{brzozowski1962canonical}'s algorithm~\cite{brzozowski1962canonical} which first determinizes (e.g., via subset construction) the reversed \ac{nfa}, then reverses the constructed \ac{dfa}, and finally determinizes it again.
This method canonizes the original \ac{nfa} without requiring any minimization steps and can outperform classic subset construction in some cases.
Since it relies only on determinization, it can make use of OTF as well and we include it in the evaluation to show the flexibility of our approach.
Finally, \Cite{van2008five} similarly considers the potential of compression ratio by generalizing subset construction; we do a comparison against their best-performing algorithm.

\paragraph{Outline}
\Cref{sec:prelminaries} introduces preliminary notation and concepts used throughout the paper.
In \Cref{sec:otf}, we discuss the general OTF framework for which \cref{sec:registry} presents two exemplary equivalence registries.
In \Cref{sec:eval}, we evaluate the performance of OTF and compare our approach with the alternatives.
\Cref{sec:fut} concludes the paper and gives an outlook on future research.

\section{Preliminaries}
\label{sec:prelminaries}

This section introduces basic notations and concepts used throughout the paper.

\subsection{Finite Automata}

Let $\Sigma$ denote a finite, non-empty set of input symbols and $\Sigma^*$ denote the set of finite \emph{words} over $\Sigma$ with $\varepsilon$ denoting the \emph{empty word}.

\begin{definition}[Finite Automaton]
  A \ac{fa} is a tuple \linebreak ${A = (S, \Sigma, \delta, S_0, F)}$ such that
  \begin{itemize}[nosep]
    \item $S$ denotes a finite, non-empty set of states,
    \item $\Sigma$ denotes a finite, non-empty set of input symbols,
    \item $\delta \subseteq S \times \Sigma \times S$ denotes a labeled transition relation,
    \item $S_0 \subseteq S$ denotes a set of initial states, and
    \item $F \subseteq S$ denotes a set of final (or accepting) states.
  \end{itemize}
\end{definition}

Let $2^X$ denote the set of subsets of a set $X$.
By abuse of notation, we may interpret $\delta$ as a function $S \times \Sigma \mapsto 2^S$ such that $\delta(s, i) = \{t \in S \mid (s, i, t) \in \delta \}$.
We generalize this view to sets of states and words such that ${\delta(Q, i) = \bigcup_{q \in Q} \delta(q, i)}$ and ${\delta(s, \varepsilon) = s, \delta(s, iw) = \delta(\delta(s, i), w)}$ for ${Q \subseteq S, s \in S, i \in \Sigma, w \in \Sigma^*}$.
In the following, we may refer to a set of states $Q$ as a \emph{metastate}.

A word $w$ is \emph{accepted} by a state ${s \in S}$ iff ${\delta(s, w) \cap F \neq \emptyset}$.
The \emph{language} of a state $s \in S$, $L(s)$, is defined as set of accepted words of $s$.
The language of a metastate $Q$ is defined as ${L(Q) = \bigcup_{q \in Q} L(q)}$.
The language of an \ac{fa} $A$ is defined as ${L(A) = L(S_0)}$.
Two \acp{fa}, $A$ and $B$, are said to be \emph{language-equivalent} iff ${L(A) = L(B)}$.
We call $A$ a \emph{deterministic} finite automaton (\acs{dfa}) iff ${\vert S_0 \vert = 1}$ and ${\vert \delta(s, i) \vert \leq 1}$ for all ${s \in S, i \in \Sigma}$.
Otherwise, $A$ may be referred to as a \emph{nondeterministic} finite automaton (\acs{nfa}).

\subsection{Simulation and Bisimulation}
\label{subsec:simulation}
For reasoning about the behavioral relationships between states of an \ac{fa}, there exist various notions of equivalence.
A popular one is (bi-) simulation~\cite{milner89}.

\begin{definition}[Simulation]
  Let $A$ be an \ac{fa}.
  A binary relation $R \subseteq S \times S$ is called a \emph{simulation relation} for $A$ iff ${(x,y) \in R}$ implies
  \begin{itemize}
    \item $x \in F \Rightarrow y \in F$, and
    \item $\forall i \in \Sigma\colon (x, i, x') \in \delta \Rightarrow (\exists y' \in S\colon (y, i, y') \in \delta \land (x', y') \in R)$.
  \end{itemize}
\end{definition}

\begin{definition}[Bisimulation]
  Let $A$ be an \ac{fa}.
  A binary relation $R \subseteq S \times S$ is called a \emph{bisimulation relation} for $A$ iff $R$ and its inverse $R^{-1}$ are simulation relations for $A$.
\end{definition}

Note that bisimulation is stricter than simulation equivalence, as it requires the inverse of the \emph{same} relation to also be a simulation.

The coarsest simulation and bisimulation relations, known as \emph{similarity} and \emph{bisimilarity}, can be computed in $\mathcal{O}(|\delta| \cdot |S|)$ time~\cite{ranzato2007new} and $\mathcal{O}(|\delta| \cdot \log |S|)$ time~\cite{valmari2009bisimilarity}, respectively.
It is well known that simulation implies finite trace (i.e., language) inclusion~\cite{clemente2019efficient, glabbeek1990linear, van2008five} and bisimulation implies finite trace (i.e., language) equivalence~\cite{glabbeek1990linear}.
Thus, bisimulation or simulation equivalences can be used to quotient \acp{nfa}, which is often done to reduce the size of \acp{nfa} before determinizing.
Note that this step is only relevant for nondeterministic systems since for \acp{dfa}, minimization and bisimulation quotienting coincide~\cite{bonchi2013checking}.

\subsection{Semilattice, Convexity, Antichain}
\label{subsec:lattice}

Let $X$ denote a finite set.
A \emph{semilattice} ${(X, \circ)}$ is an algebraic structure that consists of $X$ and a binary operation ${\circ : X \times X \rightarrow X}$ which is associative, commutative, and idempotent.
Let $(X, \circ)$ be a semilattice and ${M \subseteq X}$.
We call $(M, \circ)$ a (semi-) \emph{sub-lattice} of $(X, \circ)$ iff $(M, \circ)$ is a well-defined semilattice itself.
We call a semilattice $(M, \circ)$ \emph{convex} iff ${\{y \in X \mid \exists x, z \in M, x \preceq y \preceq z\} \subseteq M}$.
Here, $\prec$ denotes the preorder related to the respective (join/meet) operator $\circ$.
In this paper, we mainly consider variations of the well-known powerset lattice ${(2^X, \cup, \cap)}$.
In particular, ${(2^X, \cup)}$ is a join-semilattice.

Let ${L = (M, \circ)}$ be a semilattice.
The \emph{convexity closure} of $L$ is obtained by adding all missing elements to $L$ in order to make it convex.
For example, let $X = \{1, \ldots, 5\}, A = \{1,2\}, B = \{3,4\}$.
Then $L_1 = {(\{A, B, A \cup B\}, \cup)}$ is a \mbox{(semi-)} sub-lattice of $(2^X, \cup)$ and the convexity closure of $L_1$ is given by \[(\{\{1,2\}, \{3,4\}, \{1,2,3\}, \{1,2,4\}, \{1,3,4\}, \{2,3,4\}, \{1,2,3,4\}\}, \cup).\]
In particular, the convexity closure of $L_1$ is a well-defined sublattice of $(2^X, \cup)$.

Testing whether an element is contained in a convex (join-) semilattice $L$ can be efficiently implemented by testing whether it is covered from above by $L$'s greatest element and covered from below by any of $L$'s minimal elements.
Let ${L = (Y, \cup)}$ denote a (join-) semilattice for some ${Y \subseteq 2^X}$.
The \emph{greatest element} of $L$ is given by ${\bigcup L = \bigcup_{Y_i \in Y} Y_i}$.
An element ${Y_i \in Y}$ is called \emph{minimal} iff ${\nexists Y_j \in Y\colon Y_j \subset Y_i}$.
Here, $\subset$ is the related preorder of the join operator ($\cup$) without equality.

For $L_1$, the greatest element is $\{1,2,3,4\}$ and the minimal elements are $\{1,2\}$ and $\{3,4\}$.
The element $\{1,2,3\}$ is contained in the convexity closure of $L_1$ because ${\{1,2\} \subseteq \{1,2,3\}}$ and ${\{1,2,3\} \subseteq \{1,2,3,4\}}$.
It is worth noting that the minimal elements of $L$ form an antichain, i.e., a set of pairwise incomparable (in the sense of $\prec$) elements, and can be efficiently stored in a list-like structure~\cite{cadilhac2025data}.

\section{The OTF Framework}
\label{sec:otf}

This section presents how we integrate the OTF concept into the determinization process.
Intuitively, our approach is similar to classic subset construction with two major modifications.

First, the \ac{nfa} state space exploration may be repeatedly interrupted by a \emph{threshold} predicate.
During this interrupt, the partially constructed \ac{dfa} is minimized by an efficient algorithm such as~\cite{hopcroft1971n}.
On the one hand, this reduces the size of the intermediate \ac{dfa} model.
On the other hand, the information about equivalent \ac{dfa} states may be used to optimize the remaining exploration of the \ac{nfa} state space (see below).

Second, the mapping between \ac{nfa} metastates and \ac{dfa} states, that is established during subset construction is handled by an \emph{equivalence registry}.
Equivalence registries allow one to incorporate \emph{external} information into this mapping either by means of preprocessing steps such as (bi-)simulation relations on \ac{nfa} states or in-processing steps such as \ac{dfa} state equivalences arising from minimization.
For this, they need to support a specific set of operations.

\begin{definition}[Equivalence Registry]
  Let ${A = (S, \Sigma, \delta, S_0, F)}$ be the input \ac{nfa} of the determinization process and ${B = (S', \Sigma, \delta', S'_0, F')}$ be the resulting output \ac{dfa} with ${L(A) = L(B)}$.
  Furthermore, let ${Q \subseteq S}$ be a metastate of $A$ and ${q, q_1, q_2 \in S'}$ be states of $B$.
  An \emph{equivalence registry} is a data structure that supports the following operations:
  \begin{itemize}
    \item GET$(Q)$ returns a state $q \in S'$ such that $L(Q) = L(q)$.
    The method may also return \emph{undefined} if it is not aware of such state yet.
    \item PUT$(Q, q)$ links the behavior of metastate $Q$ to state $q$ with the precondition that $L(Q) = L(q)$.
    \item UNIFY$(q_1, q_2)$ informs the registry about the language equivalence of states $q_1$ and $q_2$, which may affect future results of the GET method.
  \end{itemize}
\end{definition}

The goal of equivalence registries is to allow for potentially more involved lookups of mapped states in order to reduce the number of explored \ac{nfa} metastates and save runtime.
For example, to replicate classic subset construction, one can implement a one-to-one equivalence registry with a hash-based data structure to run GET and PUT in (amortized) constant time and implement UNIFY as a no-op.
However, this also results in the full exploration of the \ac{nfa} state space.
Relaxing the GET operation and allowing one to look up mapped states \emph{modulo language equivalence} has huge potential for reducing the metastate space that needs to be explored, thus improving overall determinization performance.
\cref{sec:registry} looks at potential registry implementations in more detail.

We design OTF as a \emph{framework} by allowing the threshold and equivalence registry to be parameters to this process.
In practice, this allows users to construct highly customizable canonization processes tailored to their use cases.
We continue with outlining some of the overarching structure of our approach.

\subsection{Algorithmic Sketch}

\cref{alg:otf} formalizes and outlines a possible implementation.
While our full implementation includes additional technical optimizations such as batching calls or re-using buffers, we focus here on the general concepts.

\begin{algorithm}
  \caption{On-The-Fly Determinization}
  \label{alg:otf}
  \begin{algorithmic}[1]
    \Require{NFA $A = (S, \Sigma, \delta, S_0, F)$, predicate \textsc{threshold}, registry \textsc{R}}
    \Ensure{A language-equivalent DFA $A' = (S', \Sigma, \delta', S'_0, F')$}
    \State $\mathit{cntr} \gets 0$ \Comment{Initialization}
    \State $S' \gets \{\mathit{cntr}\}$
    \State $S'_0 \gets \{\mathit{cntr}\}$
    \If{$S_0 \cap F \neq \emptyset$}
        \State $F' \gets \{cntr\}$
    \Else
        \State $F' \gets \emptyset$
    \EndIf
    \State \Call{put$_R$}{$S_0$, $cntr$}
    \State $Q \gets \{S_0\}$
    \State $E' \gets \emptyset$
    \While{$Q \neq \emptyset$} \label{line:expl} \Comment{Exploration loop}
    \State $C \gets$ \Call{pop$_Q$}{$ $}
    \State $c' \gets$ \Call{get$_R$}{$C$}
    \For{$i \in \Sigma$}
        \State $N \gets \delta(C, i)$
        \State $n' \gets$ \Call{get$_R$}{$N$}
        \If{$n'$ is undefined}
            \State $\mathit{cntr} \gets \mathit{cntr} + 1$
            \State $n' \gets \mathit{cntr}$
            \State $S' \gets S' \cup \{n'\}$
            \If{$N \cap F \neq \emptyset$}
                \State $F' \gets F' \cup \{n'\}$
            \EndIf
            \State \Call{put$_R$}{$N$, $n'$}
            \State \Call{push$_Q$}{$N$}
        \EndIf
        \State $\delta'(c', i) \gets n'$
    \EndFor
    \State $E' \gets E' \cup \{c'\}$
    \If{\Call{threshold}{$A'$}} \label{line:otf} \Comment{On-the-fly minimization}
    \For{$i \in \{0, \ldots, \mathit{cntr}\}$}
        \If{$i \in F' \cap E'$}
            \State $\mathit{sig}[i] =$ \textbf{true}
        \ElsIf{$i \in S' \cap E'$}
        \State $\mathit{sig}[i] =$ \textbf{false}
        \Else
            \State $\mathit{sig}[i] = i$
        \EndIf
    \EndFor
    \For{$(s', t') \in$ \Call{minimize}{$A'$, $\mathit{sig}$}}
        \State \Call{unify$_R$}{$s'$, $t'$}
    \EndFor
    \EndIf
    \EndWhile
    \State \Return $A'$
\end{algorithmic}
\end{algorithm}

The algorithm starts by initializing a counter used to identify states of the resulting \ac{dfa} and creates the corresponding state sets (regular, initial, and final).
The equivalence registry and exploration queue%
\footnote{
We choose a last-in-first-out queue (stack) which traverses the \ac{nfa} state space depth-first.
In our experiments, this has yielded better results as cycles are found earlier, improving the detection of equivalent states.
}
$Q$ are then initialized with information about the initial \ac{nfa} metastates.
The set $E'$ tracks the fully explored states of the final \ac{dfa} $A'$.

The main exploration loop starts at \cref{line:expl}.
The \emph{current} metastate $C$ is removed from the queue and its representative $c'$ is retrieved from the registry $R$.
Every element in $Q$ has a representative in $R$, ensuring that $c'$ is always defined.
For each input symbol, the \emph{next} metastate $N$ is computed and its representative $n'$ is looked up.
If $n'$ is not found, a new state is added to $A'$ by incrementing the counter and updating the relevant variables.
The registry is then updated and $N$ is queued for future exploration.
Afterward, the respective transition can be set.
Once all outgoing transitions of $c'$ are set, $c'$ is marked as explored.

In \cref{line:otf}, the \textsc{threshold} predicate determines if minimization should occur.
To do so, the predicate may analyze various aspects of $A'$ including the number of states or transitions.
The check can also be stateful in that it only passes if a certain increase (e.g., in size) from previous invocations is met.
To perform the minimization, an associative array $\mathit{sig}$ is built, storing each state's $\mathit{signature}$ using a Boolean value for explored states' acceptance or a unique numerical identifier for unexplored states.

$A'$ is then minimized using $\mathit{sig}$ as the initial partition.
Note that $A'$ may be partial during the exploration loop but the minimization procedure may easily add an implicit sink state to ensure a correct result (e.g., when using \citeauthor{hopcroft1971n}'s algorithm~\cite{hopcroft1971n} for minimization).
For convenience, we assume that \textsc{minimize} mutates $A'$ directly, eliminating the need for explicit updates to related sets and functions.
Additionally, \textsc{minimize} returns pairs of equivalent states which are forwarded to the registry to refine future lookups (cf. \cref{sec:registry}).
The exploration loop then proceeds with the next metastate.

Once the exploration queue is empty, the final iteration of $A'$ is returned.
While $A'$ may be minimized during exploration, the threshold predicate might not trigger in the last iteration.
Thus, a final minimization is necessary for correctly canonizing $A$.
This step is omitted from \cref{alg:otf} to allow for utilizing the presented technique directly in \citeauthor{brzozowski1962canonical}'s approach as well (cf. \cref{sec:eval}) which does not require an explicit minimization.

\subsection{Termination and Correctness}

With a threshold that constantly returns \textbf{false} and a registry that implements a one-to-one mapping between metastates and representatives, \cref{alg:otf} coincides with classic subset construction.
Thus, we omit general discussions about termination and correctness, as they follow directly from the original algorithm.
Instead, we focus on how our modifications preserve these properties.

The first modification relaxes GET to return representatives of language-equivalent metastates instead of previously mapped ones.
Once classic subset construction terminates, these metastates correspond to equivalent \ac{dfa} states, which can be permuted or merged without affecting the language of the \ac{dfa}.
Allowing these transformations earlier (during lookup) only shortcuts these steps while preserving language equivalence.

The second modification introduces intermediate minimizations, where unexplored states are assigned unique blocks in the initial partition, ensuring they remain distinct until their behavior is fully determined.
Consequently, minimization never merges states with incomplete information.
Only explored states with identical acceptance and identical successors are merged, meaning that the final \ac{dfa} remains equivalent to its non-minimized counterpart.
Thus, intermediate minimizations do not alter the \ac{dfa}’s language, maintaining parity with classic subset construction.

\subsection{Complexity}

Providing a detailed analysis of OTF's complexity is a challenge as it highly depends on user-provided parameters such as the costs of the GET/PUT/UNIFY implementations, how often the  \textsc{threshold} predicate triggers, or the compression factor of the constructed \acp{dfa}.
As a result, we only want to briefly sketch the worst-case performance of OTF and refer the reader to \cref{sec:eval} for practical performance results.

The dominating factor of \ac{nfa} canonization is the state explosion problem of subset construction when determinizing the \ac{nfa}.
There exist \acp{nfa} $A$ whose language-equivalent \ac{dfa} representation requires an exponential (in the size of $A$) number of states~\cite{Sipser0086373}.
As OTF also constructs a language-equivalent \ac{dfa}, it may have to iterate its main loop (cf. \cref{line:expl}) exponentially often.
On top of that come the costs of registry management, threshold evaluation, and minimization, which -- in the worst case -- results in additional overhead compared to classic subset construction.
\Cref{sec:registry} provides more details on the complexity of the specific registries presented in this paper.

\section{Equivalence Registries}
\label{sec:registry}

The performance of OTF highly depends on the chosen equivalence registry.
Intelligently looking up and effectively unifying language-equivalent metastates can drastically mitigate the state explosion problem of classic subset construction.
In the following, we present two exemplary implementations: \emph{CCL} generalizes state equivalences obtained from intermediate minimizations via convexity closures and \emph{CCLS} further uses state-level preorders on the input \ac{nfa} (e.g., similarity) to normalize metastates before lookup.

\subsection{CCL -- Convexity Closure Lattice}
\label{subsec:ccl}
The use of convexity closures of lattices exploits a fundamental observation of \citeauthor{bonchi2013checking}~\cite{bonchi2013checking} that for two metastates $A,B$ of an \ac{nfa}, we have
\[L(A \cup B) = L(A) \cup L(B).\]
Exploiting this property further, we can conclude for ${L(A) = L(B)}$ that
\begin{align}
  L(A \cup B) = L(A) \cup L(B) = L(A) = L(B). \label{eq:ccl1}
\end{align}
This relationship is not limited to disjoint metastates.
Consider the situation where ${L(A) = L(A \cup B)}$.
Here, no state in ${B \setminus A}$ impacts the language of either $L(A)$ or $L(A \cup B)$.
In particular, for every $X$ such that ${A \subseteq X \subseteq A \cup B}$, we have
\begin{align}
  L(A) = L(X) = L(A \cup B). \label{eq:ccl2}
\end{align}
The same reasoning applies to the case of ${L(B) = L(A \cup B)}$, placing us in a scenario similar to \cref{subsec:lattice}: detecting equivalence between two metastates allows one to construct a (join-) semilattice $L$ with minimal elements $A,B$ and greatest element ${A \cup B}$.
$L$ itself forms a sub-lattice for the powerset lattice of the global metastate space.
Due to \cref{eq:ccl1,eq:ccl2}, language equivalence holds for all elements in the convexity closure of $L$.
Thus, recognizing equivalence between two metastates $A$, $B$ provides insight into the behavior of many (potentially yet unobserved) metastates.

The CCL registry maintains a mapping between traversed states of the \ac{dfa}-under-construction and lattices that cover their respective language-equivalent metastates.
Our implementation realizes the respective operations as follows.

\subsubsection{PUT$(Q, q)$}
Initially, when metastate $Q$ is linked with state $q$, a singleton lattice is stored.
Using a hash-based data structure, this can be implemented in (amortized) constant time.

\subsubsection{UNIFY$(q_1, q_2)$}
When being informed of the equivalence of two states, their associated lattices $L_1,L_2$ must be merged.
If $L_1, L_2$ are singletons, this process coincides with the example of \cref{subsec:lattice} and is implemented straightforwardly in linear time (for set union).
If either lattice has undergone previous merges, we can apply an abstracted convexity closure: since all metastates covered by $L_1, L_2$ are language-equivalent, we select their greatest elements $G_1, G_2$ as single metastate representatives to apply the same steps as before.
The greatest element of the new lattice $L'$ is given by the joined metastate ${G_1 \cup G_2}$.
Due to union being the join operator, this ensures that $L'$ covers all elements of $L_1,L_2$ from above.
The new minimal elements of $L'$ are constructed from the union of minimal elements $M_1,M_2$ of $L_1,L_2$, respectively.
Here, elements in $M_1,M_2$ may be in a subset relation which would violate their minimality in $L'$.
To tackle this issue, we filter out these elements which requires at most $\mathcal{O}(|M_1||M_2|)$ subset comparisons which are the dominating operations of this case.

\subsubsection{GET$(Q)$}
In order to look up representatives, we first check whether $Q$ is a singleton lattice via a hash-based lookup in (amortized) constant time.
If a lattice is found, we return the single representative which trivially satisfies the required language equivalence properties.
Otherwise, we iterate over the set of lattices to find one that covers $Q$ from above and below.
Recall from \cref{subsec:lattice} that the convexity closure of a lattice can be efficiently stored using its antichain of minimal elements and greatest element.
Thus, the registry first scans for eligible lattices whose greatest element is a superset of $Q$.
Then, for each candidate, the respective antichain is searched for elements that are a subset of $Q$.
If such a candidate is found, the associated representative of the lattice is returned.
By construction of the lattices and the coverage from above and below, the returned representative is guaranteed to be language-equivalent to $Q$.
If no such lattice is found, we return \emph{undefined}.

\subsubsection{Remarks}
The data structures underlying CCL (a minimal antichain plus a greatest element, cf. \cref{subsec:lattice}) are reminiscent of antichain algorithms for universality and language inclusion~\cite{doyen2010antichain,de2006antichains}, where one maintains reachable metastates modulo one-sided subsumption to avoid redundant exploration (detecting non-universality/inclusion violations early).
The role here is different: CCL uses antichains to summarize convexity-closed equivalence regions, driven by equalities obtained from intermediate minimization.

Our implementation uses simple list-based structures which in the worst-case require a quadratic (in the number of lattices and the number of their minimal elements) number of subset comparisons for the lookup.
Future work may look at more involved structures for antichains~\cite{cadilhac2025data} or inverse indices for faster subset checking~\cite{luo2015designing}.

\subsection{CCLS -- CCL with Similarity}
\label{subsec:ccls}

The CCL registry is only able to exploit knowledge about equivalent states from calls to UNIFY\@.
Thus, the performance of the CCL registry is inherently linked to the effects of minimization.
The CCLS registry refines the CCL approach by introducing an \emph{external} preprocessing step which computes the similarity relation on the original \ac{nfa} ahead of the determinization process.
This can potentially accelerate metastate equivalence detection \emph{independently} of any minimizations.

To illustrate this concept, let $a,b$ be \ac{nfa} states and let $a$ simulate $b$, i.e., ${b \preceq a}$.
Then ${L(b) \subseteq L(a)}$ which implies ${L(\{a,b\}) = L(\{a\}) \cup L(\{b\}) = L(\{a\})}$.
In metastate terms, we can replace $\{a,b\}$ with $\{a\}$ (pruning) and $\{a\}$ with $\{a,b\}$ (saturating) without affecting language equivalence.
This principle extends to supersets as well, e.g., $\{a,b,c\}$ can be pruned to $\{a,c\}$ and $\{a,c\}$ can be saturated to $\{a,b,c\}$.
We integrate this concept into the registry operations as follows.

\subsubsection{PUT$(Q, q)$}
First, we prune $Q$ into $Q_p$ and saturate $Q$ into $Q_s$, which requires linear time in the size of $Q$ if appropriate caching of simulation relations is used.
If ${Q_p \neq Q_s}$, instead of storing a singleton lattice like CCL, we directly create a multi-element lattice with $Q_p$ as a single minimal element and $Q_s$ as the greatest element.
This lattice may already cover a larger metastate space \emph{for free} and therefore reduce the search space of future lookups.
By simulation and the results of \cref{subsec:ccl}, all covered elements are language-equivalent to $q$.

\subsubsection{UNIFY$(q_1, q_2)$}
Merging lattices is implemented identical to \cref{subsec:ccl}.

\subsubsection{GET$(Q)$}
For looking up representatives, we prune $Q$ into $Q_p$ and then search for a lattice covering $Q_p$ similar to CCL\@.
The goal of this transformation is to find a representative more often than CCL because $Q_p$ may be smaller than $Q$ and therefore may be covered by smaller lattices.
Due to the language equivalence of $Q$ and $Q_p$, the retrieved representative of $Q_p$ is also a valid representative of $Q$, ensuring correctness of the result.
While pruning introduces additional linear costs, the runtime is still dominated by the quadratic lookup process.

\subsubsection{Remarks}
Simulation preorders have been used in several ways in the past.
In antichain decision procedures for inclusion/universality~\cite{doyen2010antichain,de2006antichains} and in particular in the combination of simulation and antichains~\cite{abdulla2010simulation}, simulation is used to strengthen one-sided subsumption pruning (to normalize metastates by removing simulation-dominated states) in order to reduce exploration.
In contrast,~\cite{clemente2019efficient,van2008five} also employ simulation as a normalization/reduction device that can act in both directions by removing dominated structure and, in the saturation sense, adding implied structure, albeit for different objectives (automata reduction and determinization variants, respectively).
CCLS adopts the preorder-based normalization in a different role: for each metastate $Q$ we use pruning and saturation \emph{simultaneously} as a two-sided summary, increasing PUT coverage and improving GET hit rates via the same convexity-membership criterion as in CCL\@.

Note that the effectiveness of the CCLS registry depends on the use case.
Computing the simulation preorder can be prohibitively expensive for large or dense \acp{nfa}, so the benefit of precomputing simulation relations must outweigh the cost.
On an implementation level, we preprocess the \ac{nfa}'s initial and final states following techniques from~\cite{clemente2019efficient}.
Additionally, since the CCLS registry requires similarity information at construction time, we compute it before running OTF and use it to further quotient the system via simulation equivalence.
This also allows us to avoid challenges such as cyclic subset inclusions and tie-breaking during metastate selection~\cite{van2008five}.

\section{Evaluation}
\label{sec:eval}

This section compares the performance of OTF with classic \ac{nfa} canonization algorithms.
For the evaluation, we look at two different use cases covering practical examples from a real-world application and synthetically generated systems that exhibit specific structural properties.
We continue with a brief overview of the general benchmark structure and discuss the two use cases afterward.

\subsection{Setup}
\label{subsec:setup}

\Cref{tab:config} summarizes the considered algorithms and configurations used in the evaluation.
\textbf{OTF} describes our presented canonization process using the CCL equivalence registry, whereas \textbf{OTF-S} describes the variant using the CCLS equivalence registry.
\textbf{SC} describes the canonization process via classic subset construction, whereas \textbf{SC-S} describes a variant with pruning and saturation.
Internally, we use the OTF-S implementation with a threshold that constantly returns \textbf{false}, preventing intermediate minimizations.
This closely resembles SUBSET(compress$_{\simpreorder}$) from~\cite{van2008five}.
\textbf{BRZ} describes \citeauthor{brzozowski1962canonical}'s~\cite{brzozowski1962canonical} canonization algorithm, whereas \textbf{BRZ-S} describes a variant using SC-S for the first determinization phase.
Recall that BRZ constructs canonical automata without any minimization, i.e., the second phase does not create any equivalent states.
As a result, we use plain SC for the second phase in this configuration.
\mbox{\textbf{BRZ-OTF(-S)}} describes a variant of BRZ using OTF(-S) for the first phase.

\begin{table}[tp]%
  \caption{Algorithms and configurations used in the benchmarks.}%
  \label{tab:config}%
  \centering%
  \begin{tabular*}{0.85\textwidth}{p{0.25\textwidth}p{0.1\textwidth}|>{\centering}p{0.25\textwidth}>{\centering\arraybackslash}p{0.25\textwidth}}%
    \multicolumn{2}{c}{} & \multicolumn{2}{c}{Uses intermediate minimizations?} \\
    & & No & Yes \\ \cmidrule{2-4}
    \multirow{4}{*}{Uses simulation?} & \multirow{2}{*}{No} & \textbf{SC} & \textbf{OTF} \\
    & & \textbf{BRZ} & \textbf{OTF-BRZ} \\ \cmidrule{2-4}
    & \multirow{2}{*}{Yes} & \textbf{SC-S} & \textbf{OTF-S} \\
    & & \textbf{BRZ-S} & \textbf{OTF-BRZ-S}
  \end{tabular*}%
\end{table}

The benchmarks were executed on a server with two AMD EPYC 7763 CPUs with 2TB of RAM\@.
The resources were allocated to run multiple benchmarks in parallel.
Individual runs were single-threaded and limited to 256GB of RAM\@.

\subsubsection{Parameters}
\label{subsubsec:parameters}
All configurations use Hopcroft's algorithm~\cite{hopcroft1971n} to minimize either the intermediate automata (OTF) or the final automaton (OTF, SC).
OTF and BRZ-OTF configurations use an \emph{adaptive} threshold which triggers a minimization after every $t$ calls to the \textsc{threshold} predicate (cf. \cref{line:otf}).
$t$ is initialized to $5000$ and updated after every minimization according to the formula
\begin{align*}
  t_{\mathit{new}} &= t_{\mathit{old}} \cdot \dfrac{s_{\mathit{new}}}{s_{\mathit{old}}}
\end{align*}
where $s_{\mathit{new}}$ and $s_{\mathit{old}}$ denote the number of states of the freshly and previously minimized \ac{dfa}, respectively\footnote{We initialize $s_{\mathit{old}}$ to $5000$ to prevent a division-by-zero on the first invocation.}.
This means that if the constructed \ac{dfa} does not compress well, i.e., ${s_{\mathit{new}} > s_{\mathit{old}}}$, the threshold and thus the minimization interval increases.
In contrast, when observing effective reductions in size, i.e., ${s_{\mathit{new}} < s_{\mathit{old}}}$, the threshold triggers minimizations more often to adaptively exploit this trend.
Furthermore, we limit the minimal value and the maximum increase of $t$ to $5000$ in order to better deal with extreme corner cases.

\subsubsection{Preprocessing}
To normalize the input automata, we further apply some typical redundancy elimination techniques in a preprocessing step,
trimming each automaton (i.e., removing non-accessible and non-co-accessible states) and quotienting according to the concerned configuration either by bisimulation~\cite{valmari2009bisimilarity} or simulation equivalence~\cite{clemente2019efficient,ranzato2007new}.
We found the LIGHT- and HEAVY- reduction methods from~\cite{clemente2019efficient} to be impractical in our benchmarks due to their worst-case performance on \acp{nfa}.
However, we include their preprocessing steps of modifying initial and final states to find more simulation relations.

\subsubsection{Measurements}
For our evaluation, we look at the total runtime of the canonization process (including preprocessing and final minimization), the \emph{intermediate automaton overhead}, and the number of minimizations.
The overhead describes the difference between the number of states of intermediate automata and the final \ac{dfa}.
For SC, we use the size of the determinized automaton as reference point.
For OTF and BRZ-OTF, we use the maximum size of the intermittently minimized automata as reference point.
For BRZ, we use the size of the automaton after the first determinization phase as reference point.
Furthermore, we abort a run if the runtime exceeds a previously set limit.

\subsubsection{Visualization}
We visualize the runtime and overhead via cactus plots.
Here, the x-axis shows the number of canonized systems ordered by the analyzed property, i.e., the further right, the worse the process performed.
The sweet spot is the lower right, meaning that the worst runs still performed well.
Note that cactus plots typically do not allow one to compare individual systems, as the fastest system for algorithm A may not be the fastest system for algorithm B\@.

\subsection{Use Case 1: Walnut}
\label{subsec:walnut}

For this use case, we look at a total of 52 systems that arose as intermediate calculations in Büchi-based arithmetic problems using the tool Walnut from~\cite{bell2020additive, bosma2025using, currie2023properties, fici2022properties, gabric2021, meleshko2023pseudoperiodic, mignoty2024automatic, mousavi2015mechanical, schaeffer2024first, Shallit2024, shallit2022logical, Shallit2021, Shallit2025, Shallit2025c, shallit2023automatic, shallit2024power, shallit2019circular} and personal communications with Jeffrey Shallit\@.
These were particularly intensive calculations in Walnut, some running for days~\cite{shallit2024power} and some using terabytes of memory~\cite{currie2023properties}.
Some could not be solved directly in Walnut (\cite{bosma2025using,currie2023properties,Shallit2025c} for example).
In particular, the benchmark includes systems such as \texttt{tribpseudo3} (\enquote{Open Problem 30} in~\cite{meleshko2023pseudoperiodic}) which has not been successfully canonized yet.
The minimum, median, maximum, and (rounded) mean number of states of the considered systems are 64, 9824, 60317, and 10113 whereas the minimum, median, maximum, and (rounded) mean number of input symbols are 2, 8, 1323, and 89.
For this benchmark, we set a timeout of one hour.

\subsubsection{Evaluation}

\begin{figure}[!t]%
  \includegraphics[width=\textwidth,page=1]{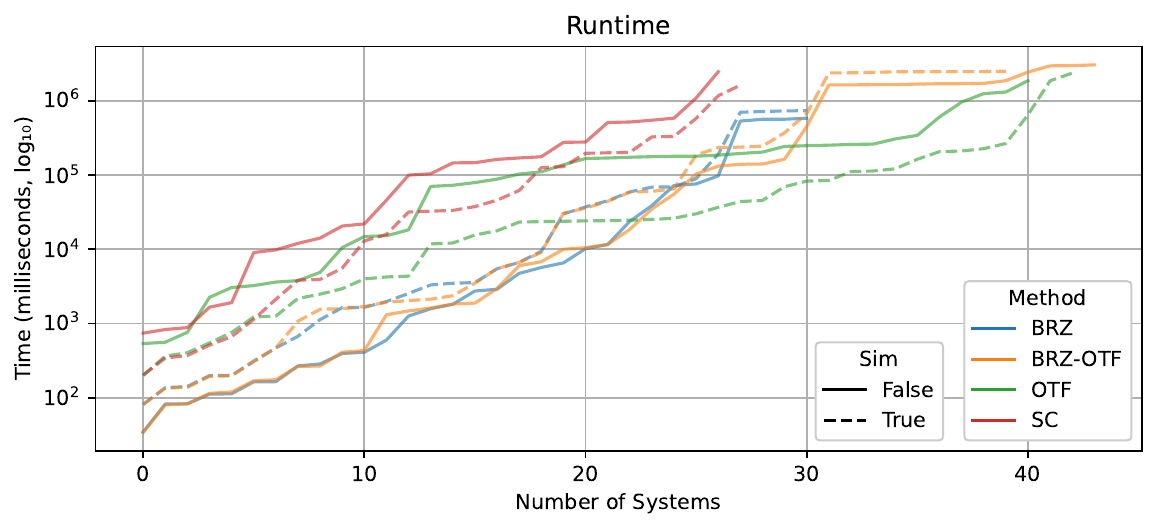}
  \includegraphics[width=\textwidth,page=2]{images/walnut}%
  \caption{Survival plots for the runtime and intermediate automaton overhead of the Walnut examples.}%
  \label{fig:walnut}%
\end{figure}%
\begin{table}[!t]%
  \caption{Number of times that the adaptive threshold triggered a minimization for the Walnut examples.}%
  \label{tab:eval-walnut}%
  \centering%
  \begin{tabular*}{\textwidth}{p{0.2\textwidth}|>{\centering}p{0.2\textwidth}>{\centering}p{0.2\textwidth}>{\centering}p{0.2\textwidth}>{\centering\arraybackslash}p{0.2\textwidth}}
    & min & median & max & average \\ \midrule
    \textbf{OTF} & 2 & 48 & 145 & 50.29 \\
    \textbf{OTF-S} & 0 & 9 & 105 & 14.93 \\
    \textbf{BRZ-OTF} & 0 & 2 & 27 & 4.61 \\
    \textbf{BRZ-OTF-S} & 0 & 1 & 17 & 3.49
  \end{tabular*}%
\end{table}%

\Cref{fig:walnut,tab:eval-walnut} show the results of the benchmark.
For the non-simulation-including configurations, we see that OTF, compared to SC, reduces the overhead of the intermediate automaton models.
While the additional work of intermediate minimizations results in somewhat similar runtimes for the best cases, its positive impact on the more complex systems is evident as OTF canonizes notably more systems within the set timeout than SC\@.

For the BRZ-based approaches, we see that the general performance on the faster systems is better, which can be explained by the reduced overhead of intermediate models.
Still, more complex systems time out in the first phase.
By compressing the intermediate models, BRZ-OTF significantly canonizes more systems compared to plain BRZ\@.

When looking at the simulation-including configurations, we see two diverging developments.
For SC and OTF, simulation boosts performance both by improving runtime and reducing overhead, sometimes to the point of not introducing any redundant states at all.
For the two BRZ-based configurations, simulation introduces a small increase in runtime and impacts the overhead only very little.
An interesting comparison can be made between OTF-S and the BRZ-OTF approaches.
While OTF-S sees an increased overhead for most systems, it yields better worst-case runtime performance than BRZ-OTF for cases that do not time out.
Here, the single-phase OTF-S beats the double-phase BRZ-OTF(-S) despite greater resource consumption.
Furthermore, it is worth noting that although OTF-S and BRZ-OTF successfully canonize a similar number of systems (44 versus 45, respectively), these are not the same systems.
Only two systems cannot be successfully canonized by either (among which is the previously mentioned \texttt{tribpseudo3}).

\Cref{tab:eval-walnut} gives additional insights into the behavior of the adaptive threshold.
The gap between plain OTF and BRZ-OTF supports the previous observations that the intermediate models of the BRZ-based approaches are comparatively small and it is likely the second phase that causes timeouts.
For both configurations, including simulation information reduces the number of minimizations, which highlights how equivalence registries enable one to incorporate external information that can further reduce the exploration space of the \acp{dfa}.

Overall, the results demonstrate the positive impact that OTF can have on the canonization process as well as its flexibility by being able to boost the performance of two different algorithmic approaches.

\subsection{Use Case 2: Random Systems with Modular Structure}
\label{subsec:ranmod}

For this use case, we look at synthetic systems in order to analyze properties of OTF at different scales.
Random models like the ones by Tabakov-Vardi~\cite{tabakov2005experimental} typically lack any structure which is what OTF is designed to exploit.
Thus, we adapt the classic Tabakov-Vardi random \ac{nfa} model by introducing structural constraints based on a modular partition of the state space.
This is similar in spirit to Tabakov's own random models with linear or grid structure~\cite{tabakov2006experimental}.

Given an automaton with states $\{0, \ldots, n-1\}$, we define an alphabet of size ${k = \max\left(1,\lfloor \sqrt{n} \rfloor\right)}$ and partition the states into $k$ disjoint classes $\{C_0, \dots, C_{k-1}\}$ such that ${q \in C_i \iff (q \bmod k) = i}$.
For each class $C_i$, we designate one initial and accepting state.
For every symbol $a \in \{0,\dots,k-1\}$, transitions on $a$ from states in class $C_i$ are directed exclusively to states in class $C_{(i+a)\bmod k}$, introducing a modular structure that couples symbols and classes.
Outgoing transitions are generated independently per source state with an expected out-degree given by the \emph{density} parameter~$\rho$ and further restricted so that each class and symbol combination maintains a roughly balanced in- and out-degree distribution.

This model preserves the expected edge density of the Tabakov-Vardi random construction while embedding a modular structure.
For the benchmark, we fix the transition density at 2 as we have observed that other densities result in similar behavior.
We look at \acp{nfa} with ${n \in \{20, 30, \ldots, 300\}}$ states and for each $n$, generate 10 automata with different random seeds each.
For this benchmark, we set a timeout of 1000 seconds.

\subsubsection{Evaluation}

\begin{figure}[!t]%
  \includegraphics[width=\textwidth,page=1]{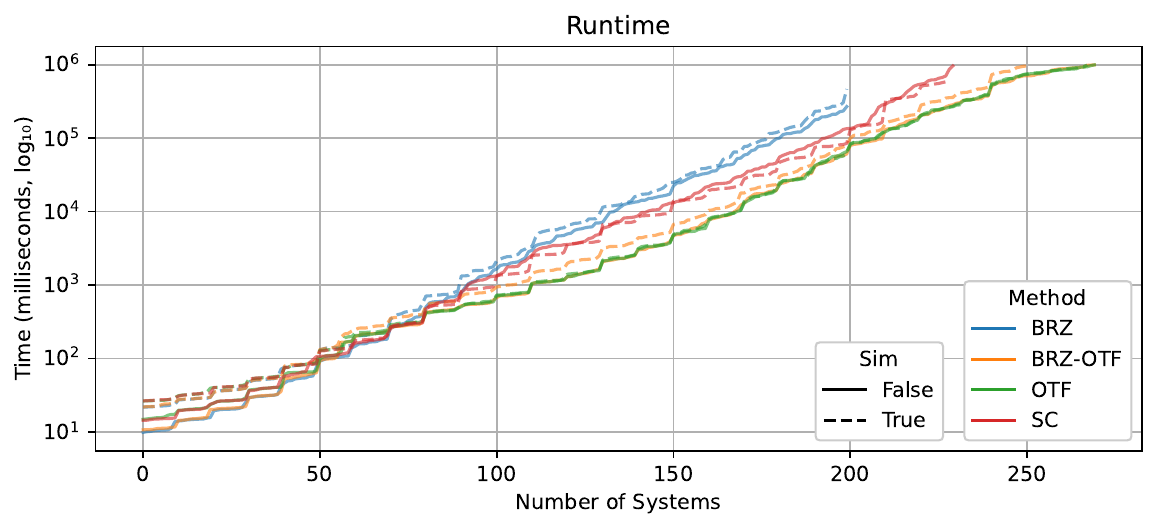}
  \includegraphics[width=\textwidth,page=2]{images/ranmod}%
  \caption{Survival plots for the runtime and intermediate automaton overhead of the random systems with modular structure.}%
  \label{fig:ranmod}%
\end{figure}%
\begin{table}[!t]%
  \caption{Number of times that the adaptive threshold triggered a minimization for the random systems with modular structure.}%
  \label{tab:eval-ranmod}%
  \centering%
  \begin{tabular*}{\textwidth}{p{0.2\textwidth}|>{\centering}p{0.2\textwidth}>{\centering}p{0.2\textwidth}>{\centering}p{0.2\textwidth}>{\centering\arraybackslash}p{0.2\textwidth}}
    & min & median & max & average \\ \midrule
    \textbf{OTF} & 0 & 6 & 129 & 25.48 \\
    \textbf{OTF-S} & 0 & 6 & 129 & 24.78 \\
    \textbf{BRZ-OTF} & 0 & 6 & 135 & 25.50 \\
    \textbf{BRZ-OTF-S} & 0 & 5 & 119 & 18.60
  \end{tabular*}%
\end{table}%

\Cref{fig:ranmod,tab:eval-ranmod} show the results of the benchmark.
Note that for the generated models, quotienting via simulation equivalence typically has no effect as states in this model will never exhibit similarity, except for some corner cases introduced by our preprocessing steps.
As a result, simulation mostly manifests as noise in the runtime and intermediate automaton overhead, most notably on single-state automata where the overhead is zero.

The step-like pattern corresponds to the increases in system size where a growing number of states correlates with a longer duration for \ac{nfa} canonization.
Similar to \cref{subsec:walnut}, we can see that the BRZ-based approaches are competitive at their fastest performances but begin to struggle in the more complex scenarios.
Only BRZ-OTF breaks this pattern and aligns with the performance of regular OTF\@.
An explanation for this is given by the automaton overhead plot which shows that after a certain size, the OTF-flavored techniques are able to exploit the structure of the modular systems and the intermediate minimizations are able to reduce the overall overhead.
This also yields improved runtimes, though less dramatically, because the equivalence classes in this model are dominated by one class with a complex antichain, leading to slower searches.

\Cref{tab:eval-ranmod} gives further insights into the behavior of the threshold.
Compared to \cref{subsec:walnut}, the differences between the configurations are rather miniscule, aligning with the previous observations that OTF itself is the key performance booster.
Again, OTF reduces overhead and improves runtime while simultaneously being flexible enough to integrate in different algorithmic concepts.

\section{Conclusion and Future Work}
\label{sec:fut}

In this paper, we present OTF -- a framework for \emph{on-the-fly} \ac{nfa} canonization.
Central to OTF are \emph{equivalence registries} which connect the determinization phase with the minimization phase to enable performance improvements.
With CCL and CCLS, we have presented two exemplary implementations of equivalence registries.
Our performance evaluation shows that for \acp{nfa} with \enquote{structure}, the discussed OTF configurations outperform classic subset construction and \citeauthor{brzozowski1962canonical}'s algorithm for complex systems while being competitive for the remaining ones.
We implement our approach in an open-source library~\cite{otf} that is based on AutomataLib~\cite{automatalib} and included in the theorem prover Walnut~\cite{walnut} since version 7.

Looking forward, there exist several directions for extension.
One obvious direction deals with exploring further implementations of equivalence registries.
\citeauthor{bonchi2013checking} establish the notion of \emph{up-to congruences}~\cite{bonchi2013checking} which capture additional relations between equivalence classes, thus further reducing exploration space.
However, in our initial experimentation, we observed a worse performance which aligns with results of others in this direction~\cite{fu2017equivalence}.
This appeared to be due to high-cardinality equivalence classes which are both slow and unlikely to match a given metastate.
A potential direction for future research is the development of hybrid approaches that apply \emph{up-to congruences} only to smaller cardinalities, or alternatively, exploit hierarchical structures.
These ideas touch on the vast field of formal concept analysis~\cite{bertet2018lattices}.

Furthermore, the presented registries of this paper may be improved.
States and transitions may be represented as binary decision diagrams~\cite{chocholaty2024mata,touzeau2019fast}, potentially combined with a semi-symbolic approach~\cite{de2006antichains}.
This may be helpful in cases where systems have large alphabets~\cite{fu2017equivalence}.
The ability to encapsulate these technical details in equivalence registries makes OTF a flexible and powerful framework for future research.

\subsubsection*{Acknowledgments}

We thank Richard Mayr for helpful advice regarding simulations and Jeffrey Shallit for discussions regarding Walnut.

\subsubsection*{Disclosure of Interests}
Markus Frohme is funded by Deutsche Forschungsgemeinschaft (DFG), Grant 528775176.
The authors are directly involved with development of the presented tool, and declare they have no financial interests.

\section*{Data-Availability Statement}
The systems used in the experiments, the full results, and an executable Docker image for reproducing the data are available in Zenodo with identifier \url{https://doi.org/10.5281/zenodo.18163403}.
For future use, the tool is maintained at \url{https://github.com/jn1z/OTF}.

\printbibliography

\end{document}